\documentclass[pra,twocolumn,altaffillsymbol,groupedaddress,floatfix,amssymb,amsmath,secnumarabic,nofootinbib]{revtex4-1}    

\usepackage[pdftex]{graphicx} 
\usepackage{dcolumn}  
\usepackage{bm}       
\usepackage[usenames,dvipsnames]{color}
\definecolor{URLCOL}{rgb}{0,0.3,0.7} 
\definecolor{LINKCOL}{rgb}{0.05,0.5,0} 
\definecolor{CITECOL}{rgb}{0.25,0,0.48} 
\usepackage[bookmarks,breaklinks,bookmarksopen,bookmarksnumbered,colorlinks,linkcolor=LINKCOL,linktocpage,citecolor=CITECOL,urlcolor=URLCOL,pdfpagemode=UseOutline,pdftex,pagebackref]{hyperref}

\usepackage{fancyhdr}
\usepackage{booktabs}  
\usepackage{dcolumn}
\newcolumntype{d}{D{.}{.}{-1}}
\usepackage{natbib}
\def\tocsecspace#1{}
\def\tocless{}

\definecolor{TITLECOL}{rgb}{0.1,0.2,0.7} 
\definecolor{CHAPCOL}{rgb}{0,0.48,0} 
\definecolor{SECOL}{rgb}{0.1,0.2,0.7} 
\definecolor{SHDCOL}{rgb}{0.4,0,0} 

\def\sec#1{\tocsecspace{20pt}\section{\textcolor{SECOL}{#1}}
\markboth{\textcolor{SECOL}{\em \bf #1}}{}
}
\def\ssec#1{\tocsecspace{10pt}\tocless\subsection{\textcolor{SSECOL}{#1}}
\markright{\textcolor{SSECOL}{#1}}
}
\def\usec#1{\section*{\textcolor{SECOL}{#1}}}

\def\coloredtitle#1{\title{\textcolor{TITLECOL}{#1}}} 
\def\coloredauthor#1{\author{\textcolor{CITECOL}{#1}}} 
\def\coltableofcontents{ 
\definecolor{SECOL}{rgb}{0.25,0,0.48} 
\definecolor{SSECOL}{rgb}{0.2,0.08,0.53} 
\tableofcontents
\definecolor{SECOL}{rgb}{0.1,0.2,0.7} 
\definecolor{SSECOL}{rgb}{0.25,0,0.48} 
}

\def\D{d}

\def\Eqsref#1{Eqs.\;\eqref{#1}}
\def\Eqref#1{Eq.\;\eqref{#1}}
\def\Secref#1{Section \ref{#1}}
\def\Figref#1{Fig.\ \ref{#1}}
\def\Ref#1{Ref.\ \cite{#1}}

\def\mnote#1{\marginpar{\footnotesize\raggedright \textcolor{red}{#1}}}

\def\revision#1{\textcolor{red}{#1}}

\definecolor{ORGCOL}{rgb}{0.8,0.65,0.1} 

\def\shd#1{\vspace{10pt}\textcolor{SHDCOL}{\bf #1:}} 

\def\bea{\begin{eqnarray}}
\def\eea{\end{eqnarray}}
\def\ben{\begin{equation}}
\def\een{\end{equation}}
\def\benu{\begin{enumerate}}
\def\enu{\end{enumerate}}

\def\bei{\begin{itemize}}
\def\eei{\end{itemize}}
\def\beit{\begin{itemize}}
\def\eit{\end{itemize}}
\def\benu{\begin{enumerate}}
\def\enu{\end{enumerate}}

\def\n{n}

\def\sss{\scriptscriptstyle\rm}

\def\g{_\gamma}

\def\hatT{{\hat T}}
\def\hatV{{\hat V}}

\def\1var{(\bx_1...\bx\N)}

\def\half{\frac{1}{2}}

\def\br{{\bf r}}

\def\bx{{x}}

\def\bj{{\bf j}}
\def\bX{{\bf X}}

\def\x{_{\sss X}}
\def\c{_{\sss C}}
\def\s{_{\sss S}}
\def\xc{{\sss XC}}

\def\N{_{\sss N}}

\def\LDA{^{\rm LDA}}
\def\GEA{^{\rm GEA}}
\def\GGA{^{\rm GGA}}

\def\unif{^{\rm unif}}

\def\ee{_{\rm ee}}

\def\ALDA{^{\rm ALDA}}

\def\up{_\uparrow}
\def\dn{_\downarrow}

\def\sph_int{ {\int d^3 r}}

\def\bemit{} 
\def\emit{} 
\def\bicol#1{\shd{#1}}

\def\bbbp{\mathcal{P}} 

\def\revision#1{#1}
\def\mnote#1{}
\def\chref#1{\textcolor{LINKCOL}{#1}}

\begin{document}

\coloredtitle{The Role of Exact Conditions in TDDFT}
\thanks{\textcolor{SHDCOL}{\footnotesize {\bf To appear in: }\\
{\em Time-dependent density functional theory}, 2$^\text{nd}$ ed., \\
edited by M. Marques, {\em et al.} (Springer, 201X).
}}
\coloredauthor{Lucas O.\ Wagner}
\email{lwagner@uci.edu}
\coloredauthor{Kieron Burke}
\affiliation{Department of Physics and Astronomy\\ 
and Department of Chemistry,\\
University of California, Irvine, CA 92697, USA}

%

\pagestyle{fancy}
\maketitle              

\coltableofcontents



\usec{Introduction}
This chapter is devoted to exact conditions in time-dependent
density functional theory.   Many conditions have been
derived for the exact ground-state density functional, and several
have played crucial roles in the
construction of popular approximations.
We believe that
the reliability of the most fundamental 
approximation of any density functional theory, the local density
approximation (LDA), is due to
the exact conditions that it satisfies.
Improved approximations should satisfy at least those conditions
that LDA satisfies, plus others.  (Which others is part of the art
of functional approximation).

In the time-dependent case, as we shall see, the 
adiabatic LDA (ALDA) plays the same role as LDA in the ground-state
case, as it satisfies many exact conditions.  But we do not have
a generally applicable improvement beyond ALDA that includes nonlocality
in time.  For TDDFT, we have a surfeit of exact conditions, but
that only makes finding those that are useful to impose an even more
demanding task.

Throughout this chapter, we give formulas for pure DFT for the sake of 
simplicity (e.g.\ $E_\xc[\n]$), but in
practice {\em spin} DFT is used (e.g.\ $E_\xc[\n\up,\n\dn]$).  
We use atomic units everywhere ($e^2 = \hbar = m_\text{\rm e} = 1$),
so energies are in units of Hartrees and distances are in Bohrs.

\sec{Review of the ground state}
In ground-state DFT, the unknown exchange-correlation energy functional, $E_\xc[\n]$, plays a crucial role.  
In fact, it is this energy that we  
typically wish to approximate with some given level of
accuracy and reliability, and {\em not} the density itself.   
Using such an approximation in a modern
Kohn--Sham ground-state DFT calculation, we can calculate the total energy of any configuration of the
nuclei of the system within the Born--Oppenheimer approximation.
In this way we can extract the bond lengths and angles 
of molecules and deduce the lowest
energy lattice structure of solids.
We can also extract forces in simulations, and vibrational frequencies
and phonons and bulk moduli.  We can 
discover response properties to both external electric fields
and magnetic fields (using spin DFT).   The accuracy of the self-consistent density is irrelevant to
most of these uses.

Given the central role of the energy, it makes sense 
to devote much effort to its study as a density functional.
Knowledge of its behavior in various limits can be crucial 
to restraining and constructing accurate
approximations, and to understanding their limitations.   
This task is greatly simplified by the fact
that the total ground-state energy satisfies the variational principle.  
Many exact conditions use this in their derivation.  

In this section we will review some of the more prominent exact conditions.
They almost all concern the energy functional, which, as mentioned above, is crucial for
good KS-DFT calculations.  We also refer the interested
reader to \Ref{perdew2003}\ for a thorough discussion.
First, we will go over some of the formal definitions in DFT.

\ssec{Basic definitions}
The XC energy as a functional of the density is written as \cite{levy1979,lieb1983}
\ben
E_\xc[\n] = \min_{\Psi\to\n}\, \langle\Psi|\, \hatT+\hatV\ee\, |\Psi\rangle - T\s[\n]-U[\n],
\een
where $\Psi$ is a correctly antisymmetrized electron wavefunction, the minimization 
of the kinetic and electron--electron repulsion energies 
is done over all such wavefunctions that yield the density $\n(\br)$, $T\s[\n]$ is the 
minimum (non-interacting) kinetic energy of a system with density $\n(\br)$, and
\ben
U[\n] = \half \int \D^3 r \int \D^3 r'\,\dfrac{\n(\br)\,\n(\br')}{|\br-\br'|}
\een
is the Hartree energy.  The XC energy is usually split into an exchange piece, $E\x$,
and a correlation piece, $E\c \equiv E_\xc - E\x$.  Exchange can be defined in 
a HF-like way in terms of the KS spin orbitals $\phi_{i\sigma}(\br)$:
\ben
E\x = -\half \sum_{i,j,\sigma}^\text{occ} \int \D^3 r \int \D^3 r'\,
\frac{\phi_{i\sigma}^*(\br)\, \phi_{j\sigma}^* (\br')\, \phi_{i\sigma} (\br')\, \phi_{j\sigma} (\br)}
{|\br-\br'|}.~~~~~ 
\label{Exdef}
\een

To perform the self-consistent calculations in the non-interacting system,
we need the functional derivative of the XC energy,  
\ben
v_\xc[\n](\br) = \dfrac{\delta E_\xc[\n]}{\delta \n(\br)}\, .\label{vxc}
\een
This is called the XC potential, 
and it is the essential  
part of the multiplicative KS potential $v\s[\n](\br)$.

\shd{Orbital dependent functionals} Some functionals are most naturally expressed 
in terms of the orbitals rather than the density.  
When varying
the orbitals of these functionals, {\em nonlocal} potentials
are obtained.  For example, varying $\phi_{i\sigma}$ in \Eqref{Exdef}
leads to the nonlocal exchange term used in HF.
There is a way to transform such orbital-dependent functionals
into local potentials as in \Eqref{vxc}.  This procedure is known
as optimized effective potential (OEP) or optimized potential method (OPM) 
and is computationally expensive \cite{kummel2008}.  Using
OEP for $E\x$ results in the exact exchange approximation (EXX) for $E_\xc$
in KS-DFT.  The Krieger, Li, and Iafrate (KLI) approximation is a way to approximately solve
EXX \cite{krieger1992a}.

\shd{Adiabatic connection} 
One can imagine smoothly connecting the interacting and non-interacting systems
by multiplying the electron--electron repulsion term 
by $\lambda$, called the coupling-constant.  Changing $\lambda$ varies the strength
of the interaction, and if we simultaneously
change the external potential to keep the density fixed, we have a family
of solutions for various interaction strengths.  
This makes all quantities (besides the density) functions of $\lambda$.
When $\lambda = 0$, one has the non-interacting KS system, and
when $\lambda = 1$, one has the fully interacting system.
The following coupling-constant relations hold.

\bemit
\bicol{XC energy $\lambda$ dependence}
Altering the coupling-constant is simply related to scaling the density:
\ben
E_\xc^\lambda[\n] = \lambda^2 E_\xc[\n_{1/\lambda}],
\een
where $\n_{1/\lambda}(\br)$ is the scaled density
\ben
\n\g(\br) \equiv \gamma^3\, \n(\gamma\br),\label{ngamma}
\een
with $\gamma = 1/\lambda$.

\bicol{Adiabatic connection formula}
By using the Hellmann--Feynman theorem, one can show:
\ben
E_\xc[\n] = \int_0^1 \D\lambda\; U_\xc^\lambda[\n]/\lambda
\label{Ecl}
\een
where $U_\xc^\lambda$ is the potential contribution to exchange-correlation 
energy ($U_\xc = V\ee - U$) at coupling-constant $\lambda$.
\emit

\ssec{Standard approximations}

Despite a plethora of approximations \cite{perdew2005}, no present-day 
approximation satisfies all the conditions mentioned in this chapter, 
as seen in tests on bulk solids 
and surfaces \cite{staroverov2004}.  With that the case, one must choose
 which conditions to impose on a given approximate form.  
Non-empirical (ab initio) approaches
attempt to fix all parameters via exact conditions \cite{perdew1996a,perdew1996b},
while good empirical approaches might include
one or two parameters that are fit to some data set \cite{becke1988b,lee1988,becke1993b}.

There are two basic flavors of approximations:  pure density functionals, which are often
designed to meet conditions on the uniform gas, and orbital-dependent functionals \cite{grabo1998}, which 
meet the finite-system conditions more naturally.  The most sophisticated approximations 
being developed today use both \cite{tao2003}. For a good discussion on what 
approximation is the right tool for the job, see \Ref{rappoport2009}.

\bemit
\bicol{LDA} The local density approximation is the bread and butter of DFT.
It is the simplest, being derived from conditions on the uniform gas 
\cite{kohn1965}.  Though it is
too inaccurate for quantum chemistry (being off by about 1\;eV or 30\;kcal/mol),
it is useful in solids and other bulk materials where the electrons almost
look like a uniform gas.  There can only be one LDA.

\bicol{GGA} The generalized gradient approximation came from trial and error
when energies were allowed to depend on the gradient of the density.  While more accurate
than the LDA (getting errors down to 5 or 6\;kcal/mol), and
thus useful for quantum chemistry applications, there is no 
uniquely-defined GGA.
BLYP is an empirical GGA that was designed to minimize the error in a particular data set.
PBE is a non-empirical GGA designed to satisfy exact conditions.

\bicol{Hybrid} Hybrids have an exchange energy which is a mixture of GGA and HF,
which attempts to get the best of both worlds:
\ben
E^\text{hyb}_\xc = E_\xc\GGA + a\,(E\x - E\x\GGA),
\een
where $E\x$ is defined in \eqref{Exdef}.  The parameter $a$ was argued to be 
0.25 for the non-empirical PBE0, but is fitted for the empirical B3LYP.

\emit

\ssec{Finite systems}
The following conditions are derived for finite systems, just
as the Hohenberg--Kohn theorem is.

\bemit
\bicol{Signs of energy components}
From the variational principle and other elementary considerations, one can deduce
\ben
E_\xc[\n] \leq 0, ~~~~~~
E\c[\n] \leq 0, ~~~~~~
E\x[\n] \leq 0.
\een

\bicol{Zero XC force and torque theorem} 
The XC potential cannot exert a net force or torque on the electrons \cite{levy1985}:
\bea
\int \D^3r\ \n(\br)\; \nabla v_\xc(\br)&=&0 \nonumber \\
\int \D^3r\ \n(\br)\;\br\times \nabla v_\xc(\br)&=& 0.
\label{gsxcforce}
\eea

\bicol{XC virial theorem} 
\ben
E_\xc[\n] + T\c[\n] = -\int \D^3r\; n(\br)\; \br\cdot \nabla v_\xc(\br),
\label{Excvir}
\een
where $T\c = T - T\s$ is the kinetic contribution to the correlation energy.
The XC virial theorem as well as the zero XC force and torque theorem are satisfied by all 
sensible approximate functionals.

\bicol{Exchange scaling} By using the scaled density \eqref{ngamma},
one can easily show
\ben
E\x[n\g]=\gamma\, E\x[\n].
\label{Exg}
\een

\begin{figure}[h]
\unitlength1in
\centering
\begin{picture}(3.3,2.4)
\put(-0.2,0){\makebox(3.3,2.4){
\includegraphics[width=3.3in]{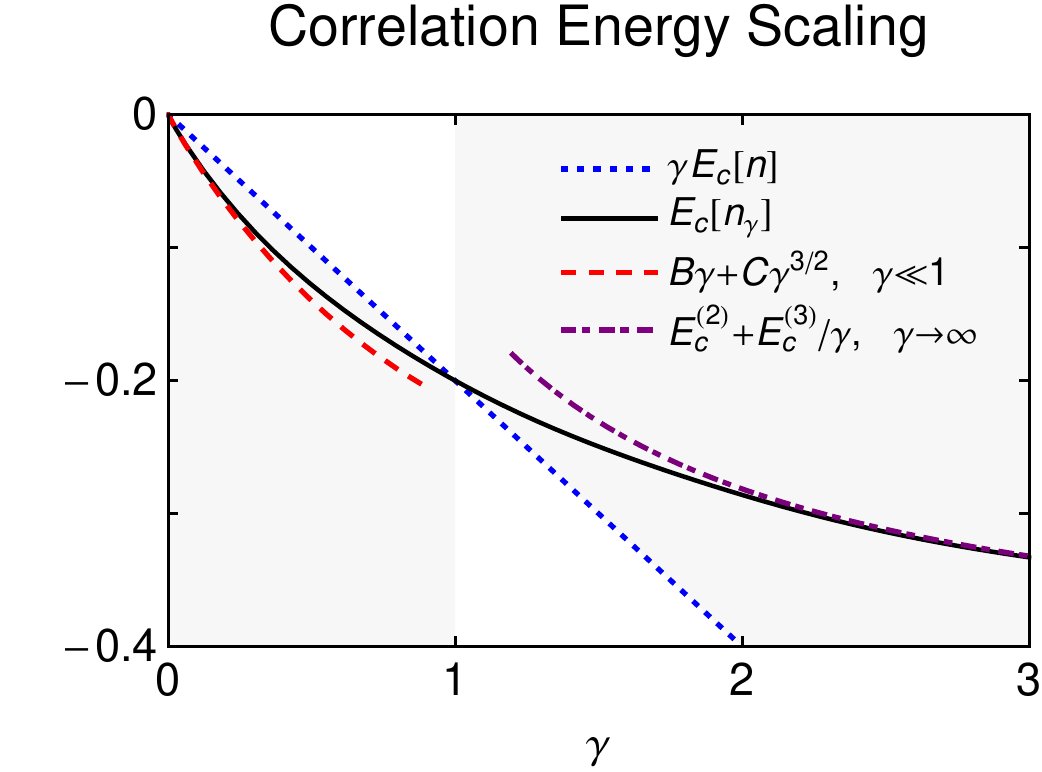}
}}
\end{picture}
\caption{Scaling of the correlation energy in ground state DFT, 
as well as the various conditions from \Eqref{Ecg}.  
The first two relations are illustrated
with the dotted line.  For $\gamma < 1$, the exact curve (solid) 
must lie below this dotted line,
and for $\gamma > 1$ the exact curve must lie above -- in both cases
within the shaded region of the graph.
The high density limit is shown with the dot-dashed
line, and the low density limit with the dashed line.
It is believed
that not only is $E\c[\n_\gamma]$ monotonic, but also its derivative
with respect to $\gamma$.  Color online.
}
\label{EcScaling}
\end{figure}

\bicol{Correlation scaling} The scaling of correlation
is less simple than exchange, and will depend on whether one
is in the high density limit ($\gamma$ large) or low density
limit ($\gamma$ small)  \cite{levy1985,seidl2000}:
\bea
E\c[\n\g] &<& \gamma\, E\c[\n]~~~~(\gamma < 1) \nonumber\\
E\c[\n\g] &>& \gamma\, E\c[\n]~~~~(\gamma > 1)
\nonumber\\
E\c[\n\g]&=&E\c^{(2)} [\n] + E\c^{(3)} [\n]/\gamma + \cdots\ \ (\gamma\to\infty)
\nonumber\\
E\c[\n\g]&=&\gamma B [\n] + \gamma^{3/2} C [\n] + \cdots\ \  (\gamma\to 0),
\label{Ecg}
\eea
where $E\c^{(2)}[\n]$, $E\c^{(3)}[\n]$, $B[\n]$, and $C[n]$ are all scale-invariant
functionals.  These conditions are depicted in \Figref{EcScaling}.  
Not all popular approximations satisfy these conditions.

\bicol{Self-interaction}  For any one-electron system \cite{perdew1981}, 
\ben
E\x[n]=-U[n],~~~E\c=0~~~~~~(N=1).\label{sione}
\een

\bicol{Lieb--Oxford bound}  For any density \cite{lieb1981}, 
\ben
E_\xc[\n] \geq 2.273\; E\x\LDA[\n].
\een
\emit

In addition to conditions on $E_\xc$,
we also know some exact conditions on the XC potential 
and the KS eigenvalues.

\bemit

\bicol{Asymptotic behavior of potential}
Far from a Coulombic system
\ben
v_\xc(\br) \to -1/r~~~~~~(r\to\infty),
\label{vxcinf}
\een
and 
\ben
\epsilon_{\rm HOMO} = - I,
\label{ehomo}
\een
where $\epsilon_{\rm HOMO}$ is the position of the
highest occupied KS molecular orbital, and $I$ the ionization
potential.
These results are intimately related to the self-interaction of one electron.
\emit

\ssec{Extended systems}\label{s:extsys}
The basic theorems of DFT are proven for {\em finite} quantum mechanical systems, with
densities that decay at large distances from the center.  Their
extension to extended systems, even those as simple as the uniform gas, requires careful thought.
For ground-state properties, one can usually take results directly to the extended limit
without change, but not always.  For example, the high-density limit 
in \Eqref{Ecg}\ 
of the correlation energy
for a finite system is violated by a uniform gas.  With these things in mind, we will now discuss
a set of conditions that involve the properties of the uniform or nearly uniform electron gas.

\bemit
\bicol{Uniform density}  When the density is uniform, $E_\xc=e_\xc\unif(n)\, {\cal V}$, where
$e_\xc\unif(n)$ is the XC energy density of a uniform electron gas of density $n$, and
${\cal V}$ is the volume.
This forms the basis of LDA.

\bicol{Slowly varying density}  For slowly varying densities, $E_\xc$ should recover the 
gradient expansion approximation (GEA):
\bea
E_\xc[\n] &=& \int \D^3r\, e\unif_\xc(\n) + \int \D^3r\, \Delta e\GEA_\xc(\n,\nabla \n) + \cdots \nonumber \\
 &=& E_\xc\LDA[\n] + \Delta E_\xc\GEA[\n] + \cdots,
\eea
where $\Delta e\GEA_\xc(\n,\nabla \n)$ is the leading correction to the
LDA XC energy density for a slowly varying electron gas \cite{langreth1980}.  However, the GEA was found to give poor results and
violate several important sum rules for the XC hole 
when applied to other systems  \cite{burke1998a}.  
Fixing those sum-rules led to the development of ab initio GGAs.  
Though important in obtaining the energy for the ground-state, the XC hole 
rules have not been used in TDDFT and therefore will not be further
discussed in this chapter.

\bicol{Linear response of uniform gas}
Another generic limit is when a weak perturbation is applied to a uniform gas,
and the resulting change in energy is given by the static response function,
$\chi(q,\omega=0)$.  This function is known from accurate
Quantum Monte Carlo calculations \cite{moroni1995}, and approximations
can be tested against it.

\emit

\sec{Overview for TDDFT}\label{s:TDview}

The time-dependent problem is more complex than the ground-state
problem, making the known exact conditions more difficult to classify.  We make the basic distinction
between general time-dependent perturbations, of arbitrary strength, and weak fields, where linear
response applies.  The former give conditions on $v_\xc[\n](\br,t)$ for
{\em all} time-dependent densities, the latter yield conditions directly on the
XC kernel, which is a functional of the ground-state density alone.
Of course, all of the former also yield conditions in the special case of weak fields.

In the time-dependent problem, we do not have the energy playing a central role.
Formally, the action plays an analogous role (see \chref{van Leeuwen ch 6}), but in practice, we
never evaluate the action in TDDFT calculations (and it is identically zero on the
real time evolution).   In TDDFT, our focus is truly the time-dependent density itself,
and so, by extension, the potential determining that density.   Thus many of our
conditions are in terms of the potential.

Most pure {\em density} functionals for the ground-state problem
produce poor approximations for the details of the potential. 
Such approximations work well
only for quantities integrated over real space, such as the energy.  Thus 
approximations that work well for ground-state energies are sometimes very poor
as adiabatic approximations in TDDFT.  
Their failure to
satisfy  \Eqref{vxcinf}\ leads to large errors in the KS energies
of higher-lying orbitals 
(for example, consider the LDA potential for Helium in Figure 3 
of \Ref{elliott2009}, which falls off exponentially rather than as $-1/r$),  
and \eqref{ehomo} is often violated by several eV.

In place of the energy, there are a variety of physical properties that
people wish to calculate.  For example, quantum chemists are most often
focused on the first few low-lying excitations, which might be crucial for
determining the photochemistry of some biomolecule.   Then the adiabatic
generalization of standard ground-state approximations is often sufficient.  At the other extreme, people who study 
matter in strong laser fields are often focused on ionization probabilities
(see \chref{Ullrich and Bandrauk chapter}), and there the violation of \Eqref{ehomo}\ makes 
explicit density approximations
too crude, and requires orbital-dependent approximations instead.

\ssec{Definitions}
In contrast to the ground-state problem, the XC potential depends not
only on the density but on the initial wavefunction $\Psi(0)$ and KS
Slater determinant $\Phi(0)$, written symbolically as $v_\xc[\n;\Psi(0),\Phi(0)](\br t)$.  
This more complicated dependence 
comes about because two different wavefunctions, which 
are chosen to have the same density for all time, 
can come from completely different external potentials,
which the XC potential accounts for.
We can get rid of this initial wavefunction dependence if we start from
a non-degenerate ground-state, where the wavefunction is a functional of the
density alone, via the Hohenberg--Kohn theorem \cite{hohenberg1964}.
These things are further discussed in \chref{Neepa's chapter}.

As the density evolves, the XC potential is determined not solely by the present
density $\n(\br,t)$, but also by the history $\n(\br,t')$ for $0 \le t' < t$.
However, it is useful to break the XC potential up into two pieces, an
{\em adiabatic} piece which only deals with the present density, and a
{\em dynamic} piece which incorporates the memory dependence:
\bea
v_\xc[\n;\Psi(0),\Phi(0)](\br t) &=&  
 v_\xc^\text{dyn}[\n;\Psi(0),\Phi(0)](\br t) \nonumber \\
&&+v_\xc^\text{adia}[\n](\br t).
\label{TDvxc}
\eea
The adiabatic piece of the potential,
\ben
v_\xc^\text{adia}[\n](\br t) = \left.\dfrac{\delta E_\xc[\n]}{\delta \n(\br)}\right|_{\n(t)},
\label{adiavxc}
\een
is the XC potential for electrons as if their instantaneous density were a ground state.
In the spirit of DFT, the dynamic piece is everything else.

In the linear response regime, small enough perturbations to the density
will continuously change the XC potential: 
\bea
\lefteqn{v_\xc[\n+\delta \n](\br t) - v_\xc[\n](\br t) =}\nonumber \\
&&\quad \quad \int \D t' \int \D^3 r'\, f_\xc[\n](\br,\br';t,t')\, \delta\n(\br't'),
\eea
where $f_\xc$ is the XC kernel, which can be written formally as the functional derivative:
\ben
f_\xc[\n_0](\br,\br';t,t') = \left.\dfrac{\delta v_\xc[\n](\br t)}{\delta \n(\br' t')}\right|_{\n_0}.
\een 
The evaluation at $\n_0$ reminds us that $f_\xc$ is
used for the linear response of a density variation
away from a ground-state density $\n_0$.

Like the XC potential, the kernel can also be broken down into an adiabatic piece: 
\ben
f_\xc^\text{adia}(\br,\br';t,t') 
= \left. \dfrac{\delta^2 E_\xc[\n]}{\delta \n(\br) \, 
\delta \n(\br')}\right|_{\n(t)} \delta(t - t'),
\label{fxcAA}
\een
and a dynamic piece, which includes memory and everything else.
The kernel
is often Fourier-transformed from position space in the
relative coordinate ($\br - \br'$) to momentum space
(with wave-vector ${\bf q}$), from the relative time ($t-t'$) 
to frequency ($\omega$) domain,
or both.  Some conditions are more naturally expressed in momentum space
and/or in the frequency domain.    
In the frequency domain, the adiabatic piece can be written as
\ben
f_\xc^\text{adia}(\br,\br') = \lim_{\omega\to 0}
f_\xc(\br,\br';\omega).
\een
The kernel
is discussed in more detail in \chref{Chapter 4 (TDDFT intro by Gross)}.

\ssec{Approximations}

As we go through the various exact conditions, we will discuss whether 
the simplest approximations in present use satisfy them.  
We can divide all approximations into two classes
based on whether or not the approximation neglects 
the dynamic term of \Eqref{TDvxc}; these classes are respectively
adiabatic and non-adiabatic (i.e.\ memory) approximations.  
In the adiabatic approximation, familiar ground-state functionals (such as LDA, GGA, and hybrids) 
can produce XC potentials when one uses the approximate $E_\xc$ in \Eqref{adiavxc}.  
We mention two notable adiabatic approximations now.

\bemit

\bicol{ALDA} The prototype
of all TDDFT approximations is the Adiabatic Local Density Approximation,
and it is the simplest pure density
functional.  The XC potential is as simple as can be:
\ben
v_\xc\ALDA[\n](\br t) = \left. \dfrac{\D e\unif_\xc(\n) }{\D \n}\right|_{\n(\br t)}.
\een
In linear response, the ALDA kernel is
\ben
f_\xc\ALDA(\br,\br';t,t') = \left.\dfrac{\D^2 e\unif_\xc(\n) }{\D \n^2}\right|_{\n(\br t)}
\, \delta^{3}(\br - \br')\, \delta(t - t').
\een %
Like its ground-state inspiration, ALDA
satisfies important sum rules by virtue of its simplicity, 
namely its locality in space and time.  ALDA is commonly used in many calculations, and is described further in \chref{chap 1}.

\bicol{AA}  In the `exact' adiabatic approximation, we use the
exact $E_\xc$ in \Eqref{adiavxc}.  
This approximation is the best that an adiabatic approximation can do,
unless there is some lucky cancellation of errors. 
Hessler {\em et al.}\ \cite{hessler2002} investigated AA 
applied to a time-dependent Hooke's atom system
and found large errors in the instantaneous correlation energy.  
For the double ionization of a model Helium atom, 
Thiele {\em et al.}\ \cite{thiele2008} discovered
 that non-adiabatic effects were
important only for high-frequency fields.

\emit

A key aim of today's methodological
development is to build in correlation memory effects.
Any attempt to build in memory
goes beyond the adiabatic approximation, and thus belongs in the 
non-adiabatic class of approximations.  The next three approximations 
belong to this dynamic class.

\bemit

\bicol{GK}  The Gross--Kohn approximation is simply to use 
the local frequency-dependent kernel of the uniform gas,  
\ben
f_\xc^\text{GK}(\br,\br'; \omega) = \delta^{3}(\br - \br')\, f_\xc\unif(\n(\br);\omega),
\een
where
\ben
f_\xc\unif(n;\omega) \equiv \lim_{q\to 0} f_\xc\unif(n;q,\omega)
\een
is the response of the uniform electron gas with density $n$.
GK was the first approximation 
to go beyond the adiabatic approximation, but was found to violate translational
invariance.

\bicol{VK}  The Vignale--Kohn approximation sought to
improve upon the shortcomings of GK.
The VK approximation is simply the gradient expansion
in the current density for a slowly-varying gas (see \chref{Vignale chapter}).

\bicol{XX} Exact exchange, the orbital-dependent functional,
is treated as an implicit
density functional (see \chref{K\"ummel's orbital chapter (11)}).
When treated this way, XX
has some memory for more than two unpolarized electrons.
\emit
With the exception of XX, non-adiabatic approximations are usually 
limited to the linear response regime and approximate the kernel, $f_\xc$.
There is now a major push to go beyond linear response
for non-adiabatic approximations.  The first such attempt was a
bootstrap approach of \Ref{dobson1997}.  More recent attempts
are described in \chref{Chapter 26 (Tokatly)} of the book and in \Ref{kurzweil2004}.

\sec{General conditions}
\label{s:gen}

In this section, we discuss conditions that apply no matter how strong
or how weak the time-dependent potential is. They apply to anything:  
weak fields, strong laser pulses, and everything in between.  
They apply also to the linear response regime, yielding the more
specific conditions discussed in \Secref{s:lin}.

\ssec{Adiabatic limit}

One of the simplest exact conditions in TDDFT is the adiabatic limit.
For any finite system, or an extended system with a finite gap, the
deviation from the instantaneous ground-state during a perturbation
(of arbitrary strength) can be made arbitrarily small.
This is the adiabatic theorem of quantum mechanics,
which can be proven by slowing
down the time-evolution, i.e., if the perturbation is $V(t)$, replacing
it by $V(t/\tau)$ and making $\tau$ sufficiently large. 

Similarly, as the time-dependence becomes very slow (or equivalently,
as the frequency becomes small), for such systems the 
functionals 
reduce to their ground-state counterparts:
\ben
v_\xc(\br, t) \to v_\xc[n(t)] (\br)~~~~~~~~(\tau\to\infty)
\een
where $v_\xc[n](\br)$ is the exact ground-state XC potential of
density $n(\br)$.

By definition, any adiabatic approximation satisfies this theorem, and
so does XX, by reducing to 
its ground-state analog for slow variations.  
On the other hand, if an approximation to $v_\xc(\br t)$ were devised
that was not based on ground-state DFT, this theorem can be used in reverse
to {\em define} the corresponding ground-state functional.

\ssec{Equations of motion}
\label{s:eom}

In this section, we discuss some elementary conditions that any reasonable
TDDFT approximation should
satisfy.  Because these conditions are satisfied by almost all approximations, they are best applied
to test the quality of propagation schemes.  
For a scheme that
does not automatically satisfy a given condition, then a numerical check of
its 
error provides a test of the accuracy of the solution.
A simple analog is the check of the virial theorem in ground-state DFT in 
a finite basis.

These conditions are all found via a very simple procedure.
They begin with some operator that depends only on the time-dependent
density, such as the total force on the electrons.  The equation of motion
for the operator in
both the interacting and the KS systems are written down, and subtracted.
Since the time-dependent density is the same in both systems, the difference
vanishes.  Usually, the Hartree term also separately satisfies the resulting
equation, and so can be subtracted from both sides, yielding a condition
on the XC potential alone.  This procedure is well-described in the \chref{Vignale
chapter} for the zero XC force theorem.

\shd{Zero XC force and torque} 
These are very simple conditions saying that interaction among the particles
cannot generate a net force \cite{vignale1995a,vignale1995b}:
\bea
\int \D^3r\ n(\br, t)\; \nabla v_\xc(\br, t)&=&0 \\
\int \D^3r\ n(\br, t)\;\br\times \nabla v_\xc(\br, t)&=&
\int \D^3r\ \br\times \frac{\partial \bj_\xc(\br, t)}{\partial t},\nonumber
\label{xcforce}
\eea
where $\bj_\xc(\br,t)$ is the difference between the interacting current
density and the KS current density \cite{vanleeuwen2001}.   The second condition
says that there is no net XC torque, {\em provided} the KS and true current
densities are identical.  This is not guaranteed in TDDFT (but is in TDCDFT).
The X-only KLI approximation, though incredibly accurate for 
ground state DFT, was found to violate the zero-force condition 
\cite{mundt2007}. 
This is  because KLI is not a solution to an approximate
variational problem, but instead an approximate
solution to the OEP equations.  
This means KLI also violates the virial
theorem \cite{fritsche1998}, which we describe next.

\shd{XC Power and Virial}
By applying the same methodology to the equation of motion for the Hamiltonian,
we find \cite{hessler1999}:
\ben
\int \D^3r\ \frac {\D n(\br t)}{\D t}\ v_\xc (\br t)
= \frac{\D E_\xc}{\D t}.
\label{Excdot}
\een
while another equation of motion yields the virial theorem, which intriguingly
has the exact same form as in the ground state, \Eqref{Excvir}:
\ben
-\int \D^3 r\ \n (\br t)\ \br \cdot \nabla v_\xc [\n] (\br t)=
E_\xc [\n] (t) + T\c [\n] (t).
\label{xcvirt}
\een
These conditions are so basic that they are trivially satisfied by any
reasonable approximation, including ALDA, AA, and XX.  
Thus they are more useful as detailed checks on a propagation scheme,
as mentioned earlier.
The correlation contribution to the latter is very small, and makes a very demanding
test.  But because the energy does not play the same central role as in the ground-state
problem (and the action is {\em not} simply the time-integral of the energy -- see \chref{Robert's
chapter 2}), testing the propagation scheme
is all they are used for so far.

\ssec{Self-interaction}
For any one-electron system,
\ben
v\x(\br,t)=-\int \D^3r'\; \frac{n(\br,t)}{|\br-\br'|},~~~
v\c(\br,t)=0~~~~~(N=1),
\label{one}
\een
These conditions are automatically satisfied by XX. 
These conditions are instantaneous in time,
so any adiabatic approximation that satisfies the ground-state conditions
of \Eqref{sione} will also satisfy these time-dependent conditions, e.g.\ AA.
On the other hand, LDA violates
self-interaction conditions in the ground-state, so ALDA 
also violates these conditions in TDDFT.

\ssec{Initial-state dependence}

There is a simple condition based on the principle that {\em any}
instant along a given density history can be regarded as the
initial moment \cite{maitra2002b,maitra2005a}.  This follows very naturally from the fact that 
the Schr\"odinger equation is first order in time.  When applied to both interacting
and non-interacting systems, we find:
\begin{equation}
v_\xc[n;\Psi(t'),\Phi(t')] (\br t)= v_\xc[n;\Psi(0),\Phi(0)] (\br
t) \, \, {\rm for} \, \, t> t',
\label{init}
\end{equation}
This is discussed in much detail in \chref{Neepa's chapter}.
Here we just mention that any adiabatic approximation, by virtue of its
lack of memory and lack of initial-state dependence, automatically satisfies it.  
Interestingly, although XX is instantaneous in the orbitals, it 
has memory (and so initial-state dependence) as a density functional (when applied
to more than two unpolarized electrons).

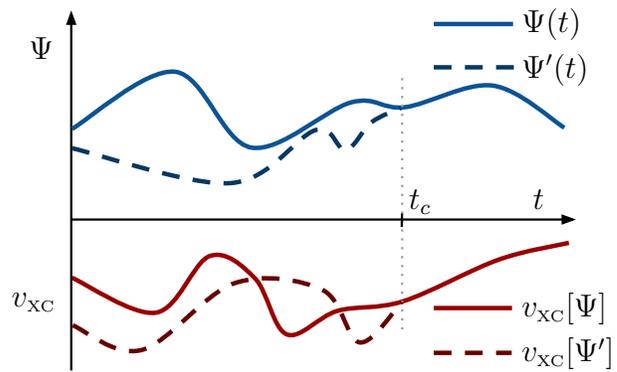
\begin{figure}[htb]
\unitlength1cm
\centering
\begin{picture}(7,5)
\put(0,0){\makebox(7,5){
	\includegraphics[width=7.1cm]{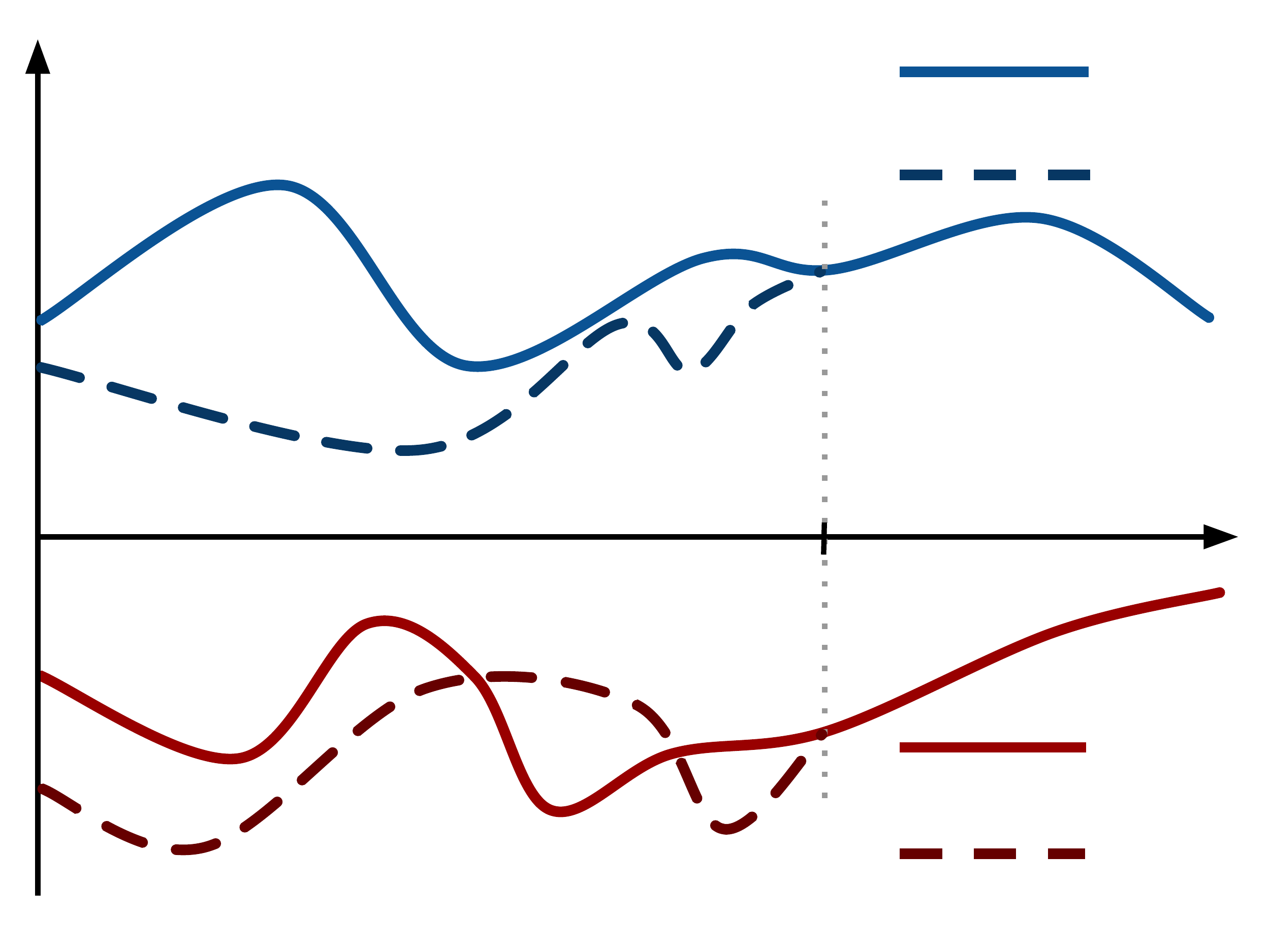}
}}
\put(6.15,4.65){\large $\Psi(t)$}
\put(6.15,4.05){\large $\Psi'(t)$}
\put(-0.4,4.3){\large $\Psi$}
\put(-0.63,1){\large $v_\xc$}
\put(6.3,2.3){\large $t$}
\put(4.65,2.3){\large $t_c$}
\put(6.15,0.87){\large $v_\xc[\Psi]$}
\put(6.15,0.27){\large $v_\xc[\Psi']$}
\end{picture}
\caption{An illustration of the condition based on initial state dependence.  The
two wavefunctions $\Psi$ and $\Psi'$ become equal at time $t_c$, and 
therefore the KS potentials must become equal then and forever after.  Color online.}
\label{InitialStateDep}
\end{figure}

This condition provides very difficult tests for any functional with memory.
Consider any two evolutions of an interacting system, whose wavefunctions $\Psi$ and $\Psi'$
become equal after some time, $t_c$.
This condition requires that the non-interacting systems have
identical XC potentials at that time and forever after, even though
they had different histories before then.  This is illustrated in
\Figref{InitialStateDep}.  An approximate functional
with memory is unlikely, in general, to produce such identical potentials.

\ssec{Coupling-constant dependence}

Because of the lack of a variational principle for the energy, 
there are no definite results for various limits, 
as in \Eqref{Ecg}, nor is there a simple extension
of the adiabatic connection formula \eqref{Ecl}, though G\"orling 
proposed an analog for time-dependent systems \cite{gorling1997}.
But there
remains a simple connection between scaling and the coupling-constant 
for the XC potential \cite{hessler1999}.
For exchange, analogous to \Eqref{Exg}, the relation is linear:
\begin{equation}
v_{\scriptscriptstyle\rm X}[n_\gamma;\Phi_\gamma(0)] ({\br} t)
= \gamma\, v_{\scriptscriptstyle\rm X}
[n;\Phi(0)]
(\gamma{\br},\gamma^2 t),
\label{vxl}
\end{equation}
where \revision{
\ben
\Phi_\gamma(0) \equiv \gamma^{3N/2}\, \Phi(\gamma \br_1,\ldots,\gamma \br_N;\, t=0)\label{phiscale}
\een
is the normalized
initial state of the Kohn--Sham system with coordinates scaled by $\gamma$}, 
and, for time-dependent densities,
\ben
n_\gamma(\br, t) \equiv \gamma^3\, n(\gamma\br, \gamma^2 t).
\een

There is no simple correlation scaling,
but we can relate the coupling-constant to scaling and find, analogous to \Eqref{Ecl}:
\bea
\lefteqn{v_{\scriptscriptstyle\rm C}^{\lambda}[n;\Psi(0),\Phi(0)] ({\br} t) =} \nonumber \\
 && \quad \quad \lambda^2 v_{\scriptscriptstyle\rm C}
[n _{1/\lambda};\Psi_{1/\lambda}(0),\Phi_{1/\lambda}(0)]
(\lambda{\br},\lambda^2 t),
\label{vcl}
\eea
where \revision{$\Psi_{1/\lambda}(0)$ is the scaled initial state of the interacting system, 
defined as in \Eqref{phiscale} and replacing $\gamma$ with $1/\lambda$.}
For finite systems, it seems likely that taking the limit $\lambda\to 0$ makes the exchange term dominant
(just as in the ground-state) \cite{hessler2002}, but this has yet to be proven.

\ssec{Translational invariance}
\label{s:TI}

Consider a rigid boost $\bX(t)$ of a system starting in
its ground state at $t=0$, with $\bX(0)=\D\bX/\D t(0)=0$.
Then
the exchange-correlation potential of the
boosted density will be that of the unboosted density,
evaluated at the boosted point, i.e.,
\ben
\label{gti}
v_\xc [n'] (\br,t) = v_\xc [n] (\br-\bX(t),t),
\een
where  $n'(\br,t) = n(\br-\bX(t),t).$
This condition is universally valid \cite{vignale1995a}. The GK
approximation was found to violate this condition, which spurred on the development of the VK approximation.

\sec{Linear response}\label{s:lin}

In the special case of linear response, all exchange-correlation information is
contained in the kernel $f_\xc$.  Linear response 
is utilized in the great majority of TDDFT calculations, and Strubbe
thoroughly discusses the methods involved in \chref{Chapter 7}.
As explained in \chref{Chapter 24 (Martin Head-Gordon)}
and \Ref{elliott2009}, the chief use of linear response has been 
to extract electronic excitations.
In this section, we shall discuss the exact conditions that pertain to $f_\xc$,
regardless of how it is employed.

\ssec{Consequences of general conditions}
Each of the conditions listed below for $f_\xc$ can be derived from a general condition
in \Secref{s:gen}.

\shd{Adiabatic limit}
For any finite system, the exact kernel satisfies:
\ben
\lim_{\omega\to 0}f_\xc(\br,\br';\omega) = \frac{\delta^2 E_\xc[n]}{\delta n(\br)\delta n(\br')}
\een
where $E_\xc$ is the exact XC energy.  Obviously, any adiabatic
functional satisfies this, with its corresponding ground-state approximation
on the right.

\shd{Zero force and torque}
The exact conditions on the potential of \Secref{s:eom}\
also yield conditions on $f_\xc$, when applied to an infinitesimal
perturbation (see \chref{Vignale chapter}).  Taking functional derivatives of
\Eqref{xcforce}\ yields
\ben
\int \D^3 r\ n(\br)\ \nabla f_\xc(\br,\br';\omega) =
-\nabla' v_\xc (\br')
\een
and
\ben
\int \D^3 r\  n(\br)\ \br\times\nabla f_\xc(\br,\br';\omega) =
-\br'\times\nabla' v_\xc (\br'),
\een
the latter assuming no XC transverse currents.
Again, these are satisfied by ground-state DFT with the static
XC kernel, so they are automatically satisfied by any adiabatic
approximation.  Similarly, in the absence of correlation, they
hold for XX.
The general conditions employing energies, \Eqsref{Excdot} and \eqref{xcvirt}, 
do not yield simple conditions for the kernel, because the functional
derivative of the exact time-dependent
XC energy is not the XC potential.

\shd{Self-interaction error}
For one electron, functional differentiation of \Eqref{one}\ yields:
\ben
f\x(\br,\br';\omega) = -1/|\br-\br'|,~~~~~f\c(\br,\br';\omega)=0~~~~~~(N=1).
\een
These conditions are trivially satisfied by XX, but violated
by the density functionals ALDA, GK, and VK. 

\shd{Initial-state dependence}
The initial-state condition, \Eqref{init}, leads to very interesting restrictions
on $f_\xc$ for arbitrary densities.  But the information is given
in terms of initial-state dependence, which is very difficult to
find.

\shd{Coupling-constant dependence}
The exchange kernel scales linearly with coordinates,
as found by differentiating \Eqref{vxl}:
\ben
f\x[n_\gamma](\br,\br',\omega)=
\gamma\, f\x[n](\gamma\br,\gamma\br',\omega/\gamma^2).
\een
A functional derivative and Fourier-transform of \Eqref{vcl}\ 
yields \cite{lein2000b} \mnote{$f\c[n]$ used to be $n_0$}
\ben
f\c^\lambda[n](\br,\br',\omega)=
\lambda^2 f\c[n_{1/\lambda}](\lambda\br,\lambda\br',\omega/\lambda^2).
\een
These conditions are trivial for XX.  They can be used to test the
derivations of correlation approximations in cases where the coupling-constant
dependence can be easily deduced.  More often, they can be used to
{\em generate} the coupling-constant dependence when needed, such
as in the adiabatic connection formula of \Eqref{Ecl}.

A similar condition has also been derived for the coupling-constant
dependence of the vector potential in TDCDFT \cite{dion2005}.

\ssec{Properties of the kernel}\label{s:kernprop}
The kernel has many additional properties that come from its
definition and other physical considerations.

\shd{Symmetry}
Because the susceptibility is symmetric, so must also be the kernel:
\ben
\label{fxcomegatr}
f_\xc({\br},{\br'};\omega)
= f_\xc({\br'},{\br};\omega) \,.
\een
This innocuous looking condition is satisfied by any adiabatic approximation
by virtue of the kernel being the second derivative of an energy, and is
obviously satisfied by XX.  

\shd{Kramers--Kronig}\label{s:KK}
The kernel
$f_\xc(\br,\br',\omega)$ is an analytic function of $\omega$ in the
upper half of the complex $\omega$-plane and approaches a real function
$f_\xc(\br,\br';\infty)$ for $\omega \to \infty$.
Therefore, defining the
function
\ben
\Delta f_\xc (\br,\br',\omega)=
f_\xc(\br,\br',\omega) - f_\xc(\br,\br';\infty),
\een
we find
\begin{equation} \label{ReKK}
\Re \Delta f_\xc(\br,\br',\omega) =  \bbbp \int \!\! \,\frac{\D
\omega'}{\pi} \frac{\Im f_\xc(\br,\br',\omega')}{\omega'- \omega}
\end{equation}
and
\begin{equation} \label{ImKK}
\hspace{6mm}\Im f_\xc(\br,\br',\omega) = -  \bbbp \int \!\! \,\frac{\D
\omega'}{\pi} \frac{\Re \Delta f_\xc(\br,\br',\omega')}
{\omega'-
\omega} .
\end{equation}
The kernel $f_\xc(\br,\br';t,t')$ is real-valued in the space and time domain,
which leads to the condition in the frequency domain,
\ben
\label{fxcomega}
f_\xc({\br},{\br'};\omega) =
f_\xc^{*}({\br},{\br'};-\omega) \,.
\een

The simple lesson here is that 
any adiabatic kernel (no frequency dependence) is
purely real, and any kernel with memory has an imaginary part 
in the frequency domain (or else is not
sensible).  Many of the failures of current TDDFT approximations,
e.g.\ the fundamental gap of solids,
are linked to the lack of an imaginary part of the kernel 
\cite{giuliani2005}.  Because adiabatic approximations produce real kernels,
we see that memory is required to produce complex kernels.  
Hellgren {\em et al.} \cite{hellgren2009} showed that XX has a complex kernel,
since it has frequency-dependence (for more than 2 electrons).
Both GK and VK have complex kernels satisfying
the Kramers--Kronig conditions.

\shd{Adiabatic connection}\label{s:adia}
A beautiful condition on the exact XC kernel is given simply by the
adiabatic connection formula for the ground-state  correlation energy:
\bea
E\c &=& -\half\int \D^3r \int \D^3 r'\;v\ee(\br-\br') \int_0^\infty 
\frac{\D\omega}{\pi} \times \nonumber \\
&&\quad ~ \int_0^1 \D \lambda\ \Im
\left[{\chi^\lambda(\br,\br';\omega)-\chi\s(\br,\br';\omega)}\right].
 \label{adaEc}
\eea
Combined with the Dyson-like equation of \chref{Chapter 1} for $\chi^\lambda$
as a function of $\chi\s$ and $f_\xc$,
this is being used to generate  new and useful approximations to 
the ground-state correlation energy \cite{fuchs2002,fuchs2005}. 
Although computationally expensive, ways are being found to speed up the 
calculations \cite{eshuis2010}.

\Eqref{adaEc} provides an obvious exact condition on any approximate XC kernel
for {\em any} system.  Thus {\em every} system for which the correlation
energy is known can be used to test approximations for $f_\xc$.
Note that, e.g, using ALDA for the kernel implicit in \eqref{adaEc}
does {\em not} yield the corresponding $E_\xc\LDA$, but rather a much
more sophisticated functional \cite{lein2000b}.   Even insertion of $f\x$ yields
correlation contributions to all orders in $E\c$.  And lastly, even
the exact adiabatic approximation, $f_\xc[\n_0] (\br,\br';\omega=0)$,
does not yield the exact $E_\xc[\n_0]$.

\shd{Functional derivatives}
A TDDFT result ought to come from a TDDFT calculation,
but this is not always the case.  By a TDDFT calculation,
we mean the result of an evolution of the TDKS equations
of \chref{chapter 1} with some approximation for the XC potential that is
a functional of the density.  This implies that the
XC kernel should be the functional derivative of some
XC potential, which also reduces to the ground-state potential
in the adiabatic limit.  All the approximations discussed
here satisfy this rule.
But calculations that intermix kernels with potentials
in the solution of Casida's equations
violate this condition, and run the risk of violating
underlying sum-rules.

\ssec{Excited states}
The following conditions have to do with the challenges
of obtaining excited states in the linear response regime.

 \shd{Infinite lifetimes of eigenstates}
This may seem like an odd requirement.  When TDDFT is
applied to calculate a transition to an 
excited state, the frequency should be real.
This is obviously true for ALDA and exact exchange, but not so clear
when memory approximations are used.  As mentioned in \Secref{s:kernprop}, 
the Kramers--Kronig relations mean that
memory implies imaginary XC kernels, and these can yield imaginary
contributions to the transition frequencies.  Such effects
were seen in calculations using the VK for atomic transitions \cite{ullrich2004}.
Indeed, very long lifetimes were found when VK was working well,
and much shorter ones occurred when VK was failing badly.

\shd{Single-pole approximation for exchange}
This is another odd condition, in which two wrongs make something right.
Using G\"orling--Levy perturbation
theory \cite{gorling1993a}, one can calculate the exact
exchange contributions to excited state 
energies \cite{filippi1997,zhang2004a}.  To recover these
results using TDDFT, one does {\em not} simply use $f\x$, and solve
the Dyson-like equations.  Like with \Eqref{adaEc}, 
the infinite iteration yields contributions to all orders in the coupling-constant.

However, the single-pole approximation truncates this series after
one iteration, and so drops all other orders.  Thus the correct
exact exchange results are recovered in TDDFT from the SPA solution
to the linear response equations, and {\em not} by a full solution \cite{gonze1999}.
This procedure can be extended to the next order \cite{appel2003}.

\shd{Double excitations and branch cuts}
Maitra et al \cite{maitra2004,cave2004} argued that a strong $\omega$-dependence in $f_\xc$
allows double excitation solutions to Casida's equations, 
which effectively couples double excitations to single excitations.
Similarly, the second ionization of the He atom implies a branch
cut in its $f_\xc$ at the frequency needed \cite{burke2005a}.
Under limited circumstances, this frequency dependence can be estimated,
but a generalization \cite{casida2005} has been proposed.  It would be
interesting to check its compliance with the conditions listed
in this chapter.

\shd{Excitations in the adiabatic approximation}
One misleading use of linear response has been 
to test the quality of different approximations to the ground-state $E_\xc$.
For instance, Jacquemin {\em et al.}\ \cite{jacquemin2010} calculated
the excitation energies for approximate $E_\xc$ functionals 
 within adiabatic TDDFT
and compared them to experimental values.  
However, even within AA -- using the adiabatic approximation with the exact $E_\xc$ -- the
exact excitations would not be not obtained.  Thus
a good ground-state $E_\xc$ used in adiabatic linear response
will not necessarily give good excitation energies.

\shd{Scattering theory and real-time propagation}
A vastly under-appreciated exact condition for TDDFT is the equivalence of time-dependent propagation and scattering theory.
This can be particularly important in understanding the relation between
bound and continuum states.

For example, much early work in TDDFT was performed by Yabana and Bertsch \cite{yabana1996}, propagating ALDA for atoms and molecules in weak electric fields. By Fourier transformation of the time-dependent dipole moment, one can extract the photoabsorption spectrum.  The fruitfly of such calculations is benzene, with a large $\pi\to\pi^*$ transition at about 6.5 eV, accurately given by ALDA. But closer inspection shows that the LDA ionization threshold is at about 5 eV, because the LDA XC potential is not deep enough.  Thus this transition is in the LDA continuum, yet its position and area are given reasonably well by ALDA. This is no coincidence:  ALDA describes the time-dependent density and its propagation for moderate times very well.  All that has changed is the choice of complete set of states onto which to project the results!

By following this logic, Wasserman {\em et al.} \cite{wasserman2003} could capture the effect of Rydberg transitions using ALDA. However, ALDA puts many bound states in the continuum due to the exponential fall-off of the KS-LDA potential (as mentioned in \Secref{s:TDview}).  Thus the ionization potentials for the ALDA states are wrong, but the oscillator strength in the LDA continuum accurately approximates that of the true Rydberg transitions to the exact bound states. 
\revision{(However, it is {\em not} an exact condition that the KS oscillator strengths be correct, not even at the threshold where KS captures the right energy \cite{yang2009}.)}
Using a trick due to Fano \cite{fano1935}, Wasserman showed \cite{wasserman2005c} that 
the quantum defect, an excruciatingly sensitive measure of the Rydberg transition frequencies, could be extracted from ALDA.  \revision{\Ref{vanfaassen2006a} shows the accuracy of this calculation for He, Be, and Ne, whereas
\Ref{vanfaassen2006b} shows the qualitative failure of ALDA for transitions to high angular momentum eigenstates (starting at the d orbitals).}

One can go further, and even consider true continuum states.
In scattering theory, the continuum states of the $N+1$ particle
system describe how a single electron scatters from  an $N$ particle system.
Wasserman \cite{wasserman2005a} and van Faassen
\cite{vanfaassen2007} developed methods to calculate
scattering amplitudes and phase shifts based on
time-propagation within TDDFT.  With a given approximation,
one can calculate the susceptibility of an atomic anion and deduce the scattering amplitude
for an incident electron \cite{wasserman2005b}.

Both these examples \revision{(the quantum defect and scattering) can be
connected in the same framework \cite{vanfaassen2009}, and they} 
illustrate that TDDFT fundamentally concerns time-propagation.  \revision{Present-day
approximations yield promising results;}
simple approximations like ALDA often yield accurate
time-dependent densities, but their projection onto individual Kohn-Sham
eigenstates may appear far more complicated.

\begin{figure}[htb]
\centering
\unitlength1cm
\begin{picture}(7,4.5)
\put(0,0){\makebox(7,4){
\includegraphics[width=8.5cm]{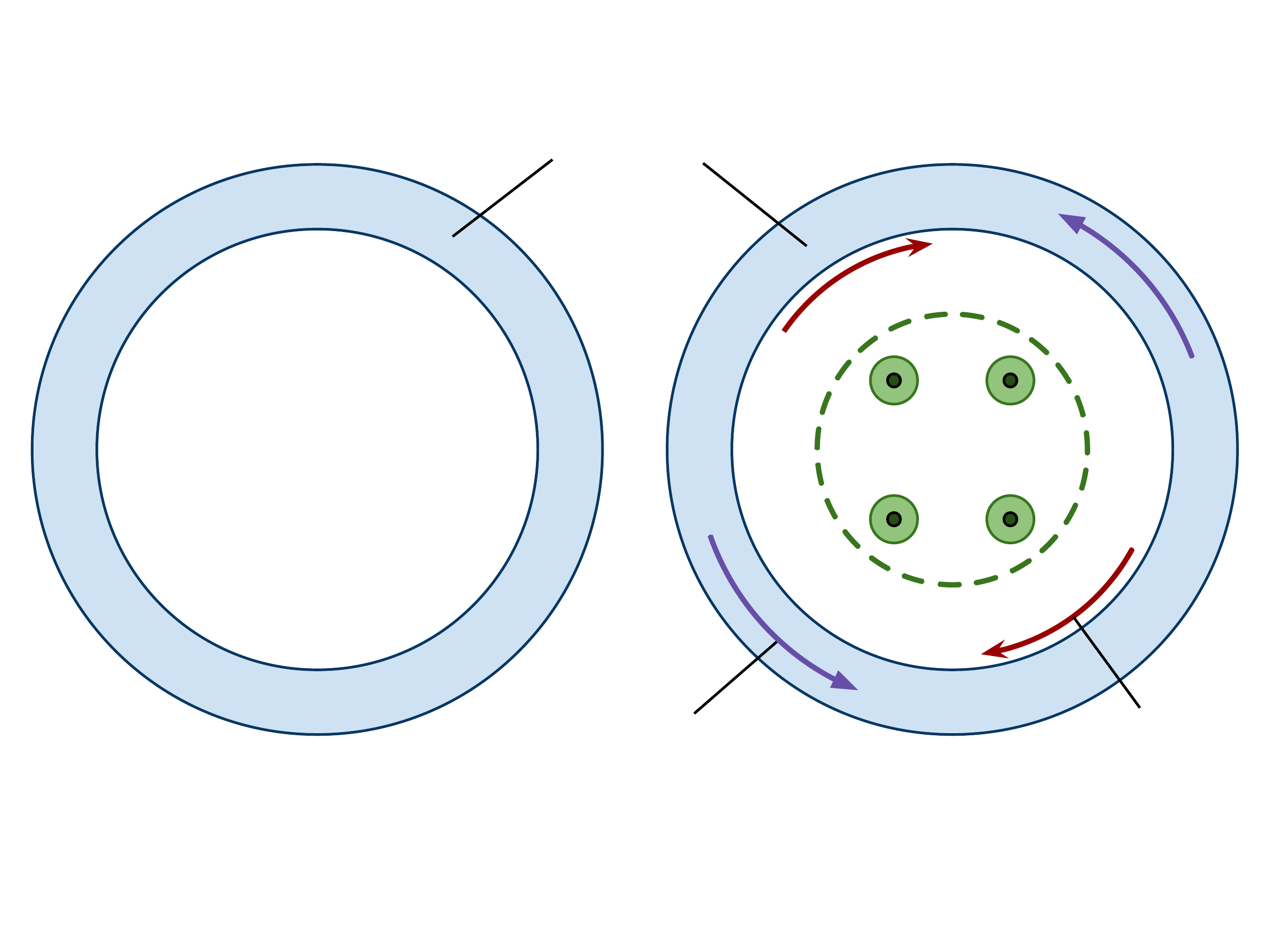}
}}
\put(-0.5,4){\large (a)}
\put(2.9,4.2){\large $\n(r,t)$}
\put(2.9,0.15){\large ${\bf j}(r,t)$}
\put(5.25,2.05){\large ${\bf B}(t)$}
\put(6.67,0.10){\large ${\bf E}(r)$}
\put(6.92,4){\large (b)}
\end{picture}
\caption{Electrons on a ring.
A magnetic field ${\bf B}(t)$ is turned on and steadily increases in (b);
the resulting electric field ${\bf E}(r)$ is uniform on a thin ring,
accelerating electrons around the ring,
producing the probability current ${\bf j}(r,t)$.  Note that in both (a)
and (b) the densities are equal.  Color online.}
\label{ring}
\end{figure}
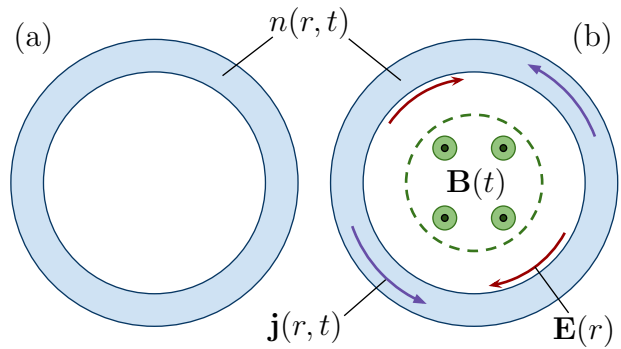

\sec{Extended systems and currents}
As mentioned in \Secref{s:extsys}, care must be taken when
extending exact ground-state DFT results to extended systems.  This is even
more so the case for TDDFT.
The first half of the RG theorem \chref{(chap 1)} provides a 
one-to-one correspondence between potentials and {\em current} densities, but
a surface condition must be invoked to produce the necessary
correspondence with densities.  
Without this condition, it can readily be seen that two
periodic systems with completely different physics can have the same density
\cite{maitra2003a},
as in \Figref{ring}.
With hindsight, this is very suggestive
that time-dependent functionals may contain a non-local dependence
on the details at a surface.  As such, they are more amenable to local approximations
in the current rather than the density.

\ssec{Gradient expansion in the current}

As discussed elsewhere (\chref{Vignale chapter}) and first pointed
out by Dobson \cite{dobson1994a}, the frequency-dependent LDA
(GK approximation) violates the translational invariance condition of \Secref{s:TI}.
One can trace this failure back to the non-locality of the XC functional
in TDDFT.  But, by going to a current formulation, everything once again
becomes reasonable.  The gradient expansion in the current, for a slowly
varying gas, was first derived by Vignale and Kohn \cite{vignale1996}, and later simplified
by Vignale, Ullrich, and Conti \cite{vignale1997}, and is discussed in much detail in the
\chref{Vignale chapter}.

For our purposes, the most important point is that, by construction, VK
satisfies translational invariance.  The frequency-dependence shuts off
(it reduces to ALDA) when the motion is a rigid translation, but turns
on when there is a true (non-translational) motion of the density \cite{vignale1996}.

Any functional with memory should recover the VK gradient expansion in
this limit, or justify why it does not.
However, the VK approximation is  {\em only} the gradient
expansion, which for the ground-state was found to
violate sum rules, as mentioned in \Secref{s:extsys}.
It is therefore likely that there exists something like a generalized gradient
approximation, which is more accurate than VK.

\ssec{Polarization of solids}
A decade ago, GGG \cite{gonze1995b} pointed out that the periodic
density in an insulating solid in an electric field
is insufficient to determine the
one-body potential, in apparent
violation of the Hohenberg--Kohn theorem \cite{hohenberg1964}.
In fact, this effect appears straightforwardly in the 
static limit of TDCDFT, and is
even estimated by calculations using the VK approximation \cite{vanfaassen2003a,maitra2003a}.
When translated back to TDDFT language, one finds a $1/q^2$ dependence
in $f_\xc$, where ${\bf q}$ is the wavevector corresponding to $\br-\br'$.
This requires $f_\xc$ to have the same degree of
nonlocality as the Hartree kernel, and this is missed by any local
or semilocal approximation, such as ALDA, but {\em is} built in
to XX \cite{kim2002a} or AA.
The need for a $1/q^2$ contribution in the optical response of solids
led to much development \cite{onida2002} for a kernel that 
allows excitons \cite{reining2002,sottile2007}.  Since the RG theorem
can be proven for solids in electric fields of nonzero $q$, one can
extract the $q \to 0$ (a constant ${\bf E}$ field)
result at the end of the calculation \cite{maitra2003a}.

\sec{Summary}

What lessons can we take away from this brief survey?

\begin{enumerate}

\item
In the ground-state theory, the total XC energy is crucial
for determining the energy of the system, and many conditions
are proven for that functional.  This is not so for TDDFT, for which
only the time-dependent density matters.  In the non-interacting system,
the KS potential, and specifically its XC component, is what counts.

\item
Explicit density functionals have poor-quality potentials,
e.g.\ LDA and GGA.  Thus successes in ground-state DFT
do not translate directly into successes in TDDFT.  
One of the greatest challenges is that the potential is
a far more sensitive functional of the density than vice versa.
Though we have enumerated many conditions on the XC potential,
it is important to determine which conditions significantly affect the
density, including those aspects of the density that are relevant to 
experimental measurements.

\item
The adiabatic approximation satisfies many exact conditions by
virtue of its lack of memory.  Inclusion of memory may lead to violations
of conditions that adiabatic approximations satisfy.   This is reminiscent
of the ground-state problem, where the gradient expansion approximation
violates several key sum rules respected by the local approximation.
Explicit imposition of those rules led to the development of generalized
gradient approximations.

\end{enumerate}

As shown in several chapters in this book, many people are presently
testing the limits of our simple approximations, and very likely,
these or other exact conditions will provide guidance on how to
go beyond them.\tocsecspace{20pt}

\section*{Acknowledgements}
We gratefully acknowledge support of 
DOE grant DE-FG02-08ER46496, and thank Stephan K\"ummel and Mark
Casida for their input, as well as Neepa Maitra for many helpful suggestions
on the manuscript.

\pagestyle{plain}
\bibliography{coolreferences}


\end{document}